\documentclass[aps,prb,reprint,twocolumn,groupedaddress,superscriptaddress,showpacs,floatfix,nobalancelastpage]{revtex4-1}
\usepackage{graphicx}
\usepackage{hyperref}
\usepackage{braket}
\usepackage{amsmath}
\usepackage{color}

\def \musr {$\mu^+$SR}

\begin{document}


\title{Quantum-critical spin dynamics in a Tomonaga-Luttinger liquid studied with muon-spin relaxation}


\author{J.~S.~M\"{o}ller}
\email{jmoeller@phys.ethz.ch}
\altaffiliation{Present address: Neutron Scattering and Magnetism, Laboratory for Solid State Physics, ETH Z\"urich, CH-8093 Z\"urich, Switzerland}
\affiliation{Department of Physics, Clarendon Laboratory, Oxford University, Parks Road, Oxford, OX1 3PU, UK}

\author{T.~Lancaster}
\affiliation{Centre for Materials Physics, Durham University, South Road, Durham, DH1 3LE, UK}

\author{S.~J.~Blundell}
\affiliation{Department of Physics, Clarendon Laboratory, Oxford University, Parks Road, Oxford, OX1 3PU, UK}

\author{F.~L.~Pratt}
\affiliation{ISIS Facility, STFC Rutherford Appleton Laboratory, Didcot OX11 0QX, UK}

\author{P.~J.~Baker}
\affiliation{ISIS Facility, STFC Rutherford Appleton Laboratory, Didcot OX11 0QX, UK}

\author{F.~Xiao}
\affiliation{Centre for Materials Physics, Durham University, South Road, Durham, DH1 3LE, UK}

\author{R.~C.~Williams}
\affiliation{Centre for Materials Physics, Durham University, South Road, Durham, DH1 3LE, UK}

\author{W.~Hayes}
\affiliation{Department of Physics, Clarendon Laboratory, Oxford University, Parks Road, Oxford, OX1 3PU, UK}

\author{M.~M.~Turnbull}
\affiliation{Carlson School of Chemistry and Biochemistry and Department of Physics, Clark University, Worcester, Massachusetts 01610, USA}

\author{C.~P.~Landee}
\affiliation{Carlson School of Chemistry and Biochemistry and Department of Physics, Clark University, Worcester, Massachusetts 01610, USA}


\date{\today}

\begin{abstract}
We demonstrate that quantum-critical spin dynamics can be probed in high magnetic fields using muon-spin relaxation (\musr). Our model system is the strong-leg spin ladder bis(2,3-dimethylpyridinium) tetrabromocuprate (DIMPY). In the gapless Tomonaga-Luttinger liquid phase we observe finite-temperature scaling of the \musr\ $1/T_1$ relaxation rate which allows us to determine the Luttinger parameter $K$. We discuss the benefits and limitations of local probes compared with inelastic neutron scattering.

\end{abstract}



\maketitle


Quantum-critical states have attracted a great amount of theoretical and experimental interest since they exhibit universal behaviour that is independent of the underlying microscopic Hamiltonian~\cite{Giamarchi,Sachdev}. Of particular interest has been the universal scaling behaviour of quantum critical phases that has so far been explored primarily using inelastic neutron scattering (INS)~\cite{Dender,Lake2005Quantum-critica,Haelg2015_1,Haelg2015_2,Povarov2015} even though many of the theoretical predictions actually concern local correlation functions which can also be explored using local probes such as nuclear magnetic resonance (NMR)~\cite{Jeong2013,Kinross2014} and muon-spin relaxation (\musr), both of which give access to an energy range effectively inaccessible to INS. \musr\ is established as a sensitive probe of magnetism and has been used to study quantum-critical spin dynamics in zero or `small' ($< 1$~T) magnetic fields\footnote{See for example, Ref.~\onlinecite{Pratt2011Nature}}. However, to the best of our knowledge, quantum-critical spin dynamics has never been explored using \musr\ in `high' ($>1$~T) applied fields, even though the majority of quantum-critical regions of interest is located at such magnetic fields. The reason for this were limitations in high-field/low-temperature capabilities of existing \musr\ instruments and the difficulty of such experiments. The commissioning of the worldwide-unique HiFi instrument at ISIS, UK~\cite{Lord_hifi} now enables \musr\ to probe spin dynamics in longitudinal fields up to 5 T at 20~mK. The longitudinal (field parallel to initial muon spin) configuration is necessary for probing spin dynamics. Here we report the observation of finite-temperature scaling of local spin correlations in the Tomonaga-Luttinger liquid phase of the strong-leg spin ladder bis(2,3-dimethylpyridinium) tetrabromocuprate (DIMPY) using \musr. This work demonstrates the feasibility of using high-longitudinal field \musr\ to study quantum-critical spin dynamics.  

Tomonaga-Luttinger liquid (TLL) theory provides a powerful, universal description of gapless interacting fermions in one dimension, equivalent to the description that Landau Fermi liquid theory provides in three dimensions~\cite{Giamarchi}. Within TLL theory the effects of interactions are contained within one single parameter: the Luttinger parameter $K$ where $K=1$ corresponds to non-interacting free fermions, $K<1$ describes repulsive interactions and $K>1$ describes attractive interactions. This parameter $K$ universally defines all correlation functions regardless of the details of the interaction potential. We note that there is an additional parameter in the Luttinger model: the velocity $u$ of the excitations, which we are not sensitive to in this work. The experimental validation of the universal finite-temperature scaling relations, predicted by TLL theory~\cite{Haldane1980}, for spin correlations in $S=1/2$ Heisenberg chains using INS~\cite{Dender,Lake2005Quantum-critica} was a particular triumph. However, Heisenberg chains, like most other experimental TLL model systems, are examples of a TLL with repulsive interactions ($K<1$)~\cite{Schulz1986,Giamarchi} and until recently a TLL with attractive interactions was only known in certain quantum Hall edge states~\cite{Grayson2007}. Spin-ladders provide unique TLL model systems since the ratio of the rung and leg exchange determines the nature of the interactions between the spinless fermions in the system, with the prototypical strong-rung ladder CuBr$_4$(C$_5$H$_{12}$N)$_2$ (BPCB) exhibiting repulsive behaviour~\cite{Klanjsek2008,Giamarchi1999} and the strong-leg ladder DIMPY exhibiting attractive interactions~\cite{Hikihara2001,Schmidiger2012,Jeong2013,Povarov2015}.

\begin{figure}[htbp]
\includegraphics[width=0.45\textwidth]{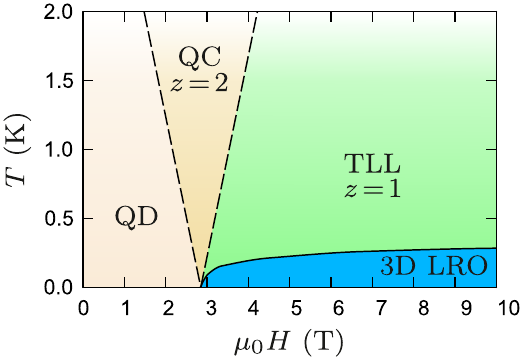}
\caption{\label{fig:pd}(colour online). Schematic phase diagram of DIMPY~\cite{Hong2010,Ninios2012,Schmidiger2012,Jeong2013}.}
\end{figure}

DIMPY is a two-leg spin ladder system with a quantum-disordered (QD) singlet ground state with gap $\Delta_0=0.32(2)$~meV in zero field~\cite{Hong2010}. By virtue of the Zeeman effect this gap can be closed by an applied field $\mu_0H_{\rm c1}=\Delta_0/g \mu_B\sim2.85$~T at a quantum critical point (QCP) with dynamic critical exponent $z=2$ (see Fig.~\ref{fig:pd})~\cite{Hong2010,Schmidiger2012}. Above $H_{\rm c1}$ and below the saturation field~\cite{White2010} $\mu_0H_{\rm c2}=30$~T the system is in the quantum critical (QC) gapless $z=1$ TLL state. The dominant exchange interactions are $J_{\rm leg}=1.42(6)$~meV and $J_{\rm rung}=0.82(2)$~meV along the ladder legs and rungs, respectively. In the gapless phase, three-dimensional long-range order (3D LRO) sets in at field-dependent critical temperatures of around 250~mK due to weak interladder interactions. Using chain mean-field (MF) theory interladder interactions were estimated~\cite{Schmidiger2012} to be $nJ_{\rm MF}'=6.5~\mu$eV, where $n$ is the number of relevant interaction pathways. 

In a \musr\ experiment spin-polarized positive muons are implanted into a sample. The experimentally-measured quantity is the decay asymmetry $A(t)$ which is proportional to the spin polarization of the muon ensemble at any one time~\cite{Blundell1999}. In this paper we will concentrate primarily on the longitudinal (spin--lattice) relaxation of the muon polarization $1/T_1$. In direct analogy with NMR, $1/T_1$ probes the local ($q$-integrated) dynamic structure factor (see Eq.~\ref{eqn:dsf})~\cite{Moriya1956}


\begin{equation}
\frac{1}{T_1}\propto\int\sum_{\rho=x,y,z} \mathcal{S}^{\rho \rho}(\mathbf{q},\omega) |_{\omega=\gamma_{\mu} B} \ d\mathbf{q},
\label{eqn:muon_t1}
\end{equation}
at the probing field $B$ and $\gamma_{\mu}$ is the muon gyromagnetic ratio. In general, \musr\ and NMR probe both longitudinal $\mathcal{S}^{zz}$ and transverse correlations $\mathcal{S}^{\perp\perp}$. Usually in a \musr\ experiment the magnetic coupling between muon and sample is primarily of dipolar nature, which leads to both transverse and longitudinal correlations being probed, while in NMR often the contact coupling is dominant, which implies that mainly transverse correlations are being probed~\cite{Yaounc}. However, the key property of the system studied here is that it is well-known that only transverse correlations exist in the TLL phase at low energies~\cite{Giamarchi1999,Hikihara2001} which facilitates the study of their properties by \musr, NMR, and non-polarized neutron scattering. For both NMR and \musr\ experiments the probing frequency corresponds to an energy scale of $~\mu$eV in fields of a few Tesla and so on the energy scale of the excitations in DIMPY and most other quantum magnets (which are on the meV scale) these effectively probe the local spin correlations as $\omega\rightarrow 0$, i.e.\ the long-time behaviour of the local spin correlations. The probing energy scale is fixed for any given field. 

Single crystal samples were grown by the method described in Ref.~\onlinecite{Shapiro2007}. Initial \musr\ experiments were performed on a mosaic of single crystals but the main \musr\ results presented here were obtained by crushing the crystals into a fine powder in order to cover a large area of the sample holder uniformly since the muon beam profile varies significantly as a function of magnetic field~\cite{Lord_hifi}. Due to the small g-factor anisotropy of the Cu$^{2+}$ ions (expected $<10$\%), the scaling functions in the TLL regime are effectively probed within a narrow range of fields but since their dependence on field in the TLL phase is relatively weak, this effect is negligible. The powder sample was mounted with vacuum grease onto a silver sample holder which was attached to the cold finger of a dilution refrigerator at the HiFi instrument, ISIS, UK. In high longitudinal fields, the relaxing asymmetry ($A_{\rm rel}$ below) is very small. The experiment therefore required high-statistics runs (40 million events) and careful attention to detector dead-time corrections by performing calibration measurements on a silver backing plate at several fields covering the field range discussed here. 

\begin{figure}[htbp]
\includegraphics[width=0.5\textwidth]{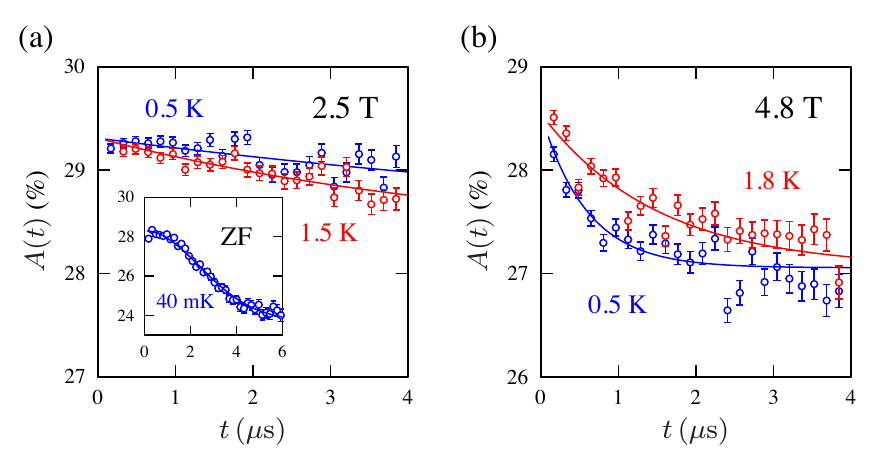}
\caption{\label{fig:asym}(colour online). Experimental muon decay asymmetry $A(t)$ (arbitrary offset). (a) Data in a longitudinal field $\mu_0H=2.5$~T on the powder sample. Inset: Data measured in zero-applied field on a mosaic of single crystals showing Gaussian relaxation due to nuclear moments. (b) Results in applied longitudinal field $\mu_0H=4.8$~T (in the TLL phase).}
\end{figure}

\begin{figure}[htbp]
\includegraphics[width=0.45\textwidth]{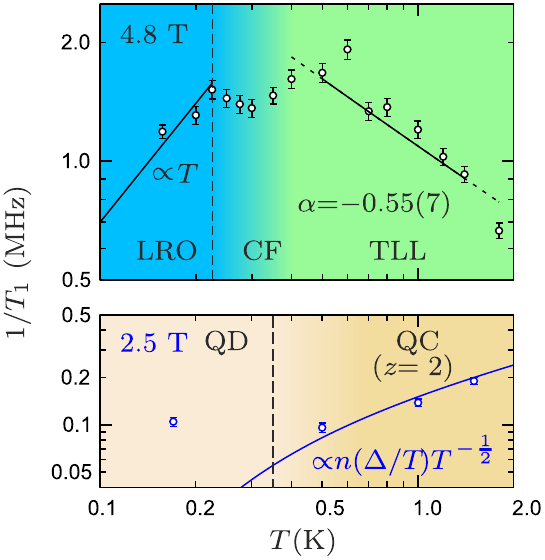}
\caption{\label{fig:scaling}(colour online). Relaxation rate $1/T_1$ at $\mu_0H=4.8$~T (top) and $\mu_0H=2.5$~T (bottom). Top: the peak around $T=225$~mK indicates long-range ordering. Above the ordering transition $1/T_1$ is first dominated by critical fluctuations before entering a regime of universal scaling for $0.4$~K~$\leq T\leq2$~K. The solid-dashed line is a fit to Eq.~\ref{eqn:DSF}. Bottom: the dashed line indicates the approximate value of the gap for $g=1.94$ at $\mu_0H=2.5$~T.}
\end{figure}

In Fig.~\ref{fig:asym} we show some of the \musr\ asymmetry data at different applied magnetic fields and temperatures. In high longitudinal fields it is not straightforward to calibrate the absolute scale of asymmetry hence the data are shown with an arbitrary offset. In zero applied field the muon spectra display a temperature-independent Gaussian relaxation down to at least 40~mK, characteristic of a relaxation due to quasistatic nuclear moments without any contribution due to electronic moments. The zero-field data therefore demonstrate the absence of a phase transition in zero-field in DIMPY down to 40~mK. At 2.5~T applied field, i.e.\ $H<H_{\rm c1}$, the relaxation is very weak at low temperatures but there is a distinct increase in relaxation rate at higher temperatures. We expect any nuclear contribution to the muon relaxation to be fully quenched at this field and therefore we are only probing the electronic spin dynamics. Fig.~\ref{fig:scaling} shows $1/T_1$ as a function of temperature at 2.5 and 4.8~T obtained by fitting the experimental asymmetry with 

\begin{equation}
A(t)=A_{\rm rel} \exp{(-t/T_1)} + A_{\rm nr},
\label{eqn:dimpy_relaxfit}
\end{equation}
where the relaxing asymmetry $A_{\rm rel}$ was kept fixed and $A_{\rm nr}$ is a field-dependent non-relaxing component\cite{supplemental}. At $\mu_0 H=2.5$~T, the system is still in the gapped singlet state and $1/T_1$ is strongly suppressed. At temperatures $T\gg nJ'=75$~mK in the $z=2$ 1D quantum critical regime  one expects~\cite{Orignac2007} $1/T_1\propto T^{-1/2}$. We observe an \emph{increase} of $1/T_1$ with increasing temperature that can be described phenomenologically by $1/T_1\propto n(\Delta/T) T^{-1/2}\sim T^{+1/2}$ [where $n(\Delta/T)$ is the Bose-Einstein occupation factor]. We note that this is similar to the behaviour observed by NMR around the $z=2$ QCP in the spin-ladder system BPCB and the gapped quantum magnet NiCl$_2$-4SC(NH$_2$)$_2$ which was related to the effect of three-dimensional interactions~\cite{Mukhopadhyay2012}. 

As we increase the field to $\mu_0H=4.8$~T and enter the TLL phase, the relaxation rate increases by approximately an order of magnitude. A temperature scan at constant field reveals a sharp rise of the relaxation rate with a peak around 225~mK followed by non-monotonic behaviour in an intermediate region between around 225 and 400~mK. Above 400~mK, $1/T_1$ exhibits power-law behaviour. We identify the peak in $1/T_1$ around 225~mK with a transition to long-range magnetic order (LRO) previously observed~\cite{Ninios2012,Schmidiger2012,Jeong2013} in DIMPY around 250~mK at $\mu_0 H=5$~T. Below the ordering temperature our data are consistent with $1/T_1\sim T$, which is expected due to the presence of a massless Goldstone mode~\cite{Giamarchi1999}. However, the paucity of data in this region prevents definitive conclusions and further work is required to study the scaling behaviour in this region. We note that the linear behaviour of $1/T_1$ in the 3D ordered state has not yet been observed by NMR~\cite{MartinK}. The region immediately above the ordering temperature $T_{\rm c}$ is dominated by thermal critical fluctuations until power law behaviour sets in around 400~mK. 
 
It has been shown that the dominant contribution to the local correlation function $\mathcal{S}(\omega)$ in a TLL at low energies is due to \emph{transverse correlations}, which is a feature that is generic to spin ladders~\cite{Giamarchi1999,Hikihara2001}. This greatly simplifies the present study as the presence of significant spectral weight in longitudinal correlations with different scaling properties at low energies would otherwise require longitudinal and transverse correlations to be studied separately. The transverse correlations take on an $\omega/T$-scaling form~\cite{Sachdev}

\begin{equation}
\frac{1}{T_1}\propto\mathcal{S}^{\perp\perp}(\omega)=(k_{\rm B}T)^\alpha F(\hbar\omega/k_{\rm B}T),
\label{eqn:DSF}
\end{equation}
where $F(\hbar\omega/k_{\rm B}T)$ is a universal function~\cite{supplemental} and $\alpha=1/2K -1$, $K$ being the Luttinger parameter~\cite{Giamarchi1999,Hikihara2001}. Given the weak temperature dependence~\cite{supplemental} of $F(\hbar\omega/k_{\rm B}T)$ in the $\omega\rightarrow 0$ limit, the temperature-dependence of $1/T_1$ approximately follows a power law 

\begin{equation}
\frac{1}{T_1}\propto T^\alpha=T^{1/2K -1}.
\label{eqn:power}
\end{equation}

Let us now consider the appropriate fitting range for extracting the TLL parameter $K$. We argued that, at our probing field $\mu_0 H=4.8$~T, thermal critical fluctuations are negligible above $\sim 0.4~$K. Furthermore the TLL model requires: (i) a linear dispersion relation and (ii) an infinitely deep Fermi sea. (i) Is found to hold up to at least 1~meV$=11.6$~K on the basis of previous INS data~\cite{Schmidiger2013_2,Povarov2015}. (ii) At $\mu_0H=4.8$~T the depth of the Fermi sea $\Delta_{\rm F}=g\mu_{\rm B} \mu_0(H-H_{\rm c1})=0.22~{\rm meV}=2.5$~K with $g=1.94$ and $\mu_0H_{\rm c1}=2.85$~T~\cite{Schmidiger2012}. Hence a fitting range extending from 0.4--1.8~K is justified within the TLL framework. To further exclude any bias due to a particular fitting range, the data were fitted four times: over the full range 0.4--1.8~K, excluding either end point, and excluding both end points. The final result is an average weighted by the inverse squared statistical errors. This method is similar to the shrinking-window method often used for extracting critical exponents. Following this procedure, fits to the full-scaling function Eq.~\ref{eqn:DSF} and the power law approximation Eq.~\ref{eqn:power} yield $\alpha=-0.55(7)$ [$K=1.10(13)$] and $\alpha=-0.62(5)$ [$K=1.33(10)$], respectively. The obtained TLL parameter $K>1$ indicates \emph{attractive} interactions between the spinless fermions in the TLL. 

\begin{figure}[htbp]
\includegraphics[width=0.5\textwidth]{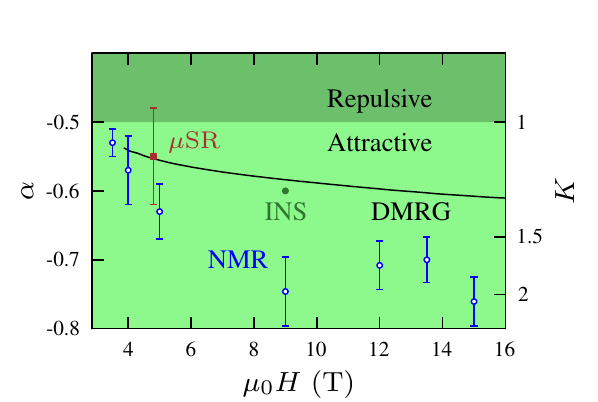}
\caption{\label{fig:alpha}(colour online). Scaling exponent $\alpha$ and corresponding Luttinger parameter $K$ measured by $^1$H NMR~\cite{Jeong2013}, INS~\cite{Povarov2015}, \musr\ (this work), and predicted by DMRG calculations~\cite{Schmidiger2012}. }
\end{figure}

Fig.~\ref{fig:alpha} shows a comparison of our \musr\ results for the Luttinger parameter with the previous experimental results based on NMR~\cite{Jeong2013} and inelastic neutron scattering~\cite{Povarov2015}. Also shown is the dependence of $K$ on the applied field based on density-matrix renormalization group (DMRG) calculations~\cite{Schmidiger2012}. We note that the scaling function $F(\hbar\omega/k_{\rm B}T)$ does have a finite temperature dependence on the \musr\ energy scales~\cite{supplemental}. Hence we expect that parameters extracted by \musr\ from Eq.~\ref{eqn:DSF} to be more accurate than those extracted from the approximation in Eq.~\ref{eqn:power}. This is consistent with the observation that the \musr\ estimates of $K$ using the former show better agreement with the DMRG calculations. NMR provides values of $\alpha$ that are lower, and correspondingly values of $K$ that are larger, than predicted by DMRG. Though only available at a single field, the \musr\ results seem to offer a somewhat better agreement with DMRG although we acknowledge that the \musr\ and NMR error bars at 4.8~T and 5~T, respectively, overlap. Further to the discussion already presented in Ref.~\onlinecite{Jeong2013} about the quantitative disagreement at higher fields between NMR and DMRG, we believe that there are two contributing factors that have not been considered so far: (i) The NMR data were analyzed using the approximation in Eq.~\ref{eqn:power}. Since $F(\hbar\omega/k_{\rm B}T)$ decreases as a function of temperature even in the applicable low-energy limit~\cite{supplemental} but is assumed constant in Eq.~\ref{eqn:power}, this leads to a fitted exponent in the power-law that is somewhat too negative. We have digitized and re-analyzed some of the published NMR data~\cite{Jeong2013} and find that this effect accounts for only approximately 1-2\% of the overestimate of $|\alpha|$. As the probing energy scales of NMR for a given field are lower than those of \musr\ by the ratio of muon and proton gyromagnetic ratios $\gamma_{\mu}/\gamma_{\rm p}\approx3.18$, the temperature-dependence of $F(\hbar\omega/k_{\rm B}T)$ is also weaker for NMR~\cite{supplemental} by approximately this factor. Hence Eq.~\ref{eqn:power} is a better approximation when analyzing NMR data than for \musr\ data. (ii) The lower energy scales probed by NMR lead to another problem: 3D interactions in DIMPY are accurately known $nJ'=6.5~\mu$eV~\cite{Schmidiger2012}. The proton NMR probing energy scale at $\mu_0 H=5$~T is $\hbar\omega=0.88~\mu$eV compared to $\hbar\omega=2.7~\mu$eV for \musr\ at $\mu_0H=4.8$~T. Therefore, any perturbing effects due to 3D interactions will be more pronounced at any given field for NMR than for \musr. 

\begin{figure}[tbp]
\includegraphics[width=0.5\textwidth]{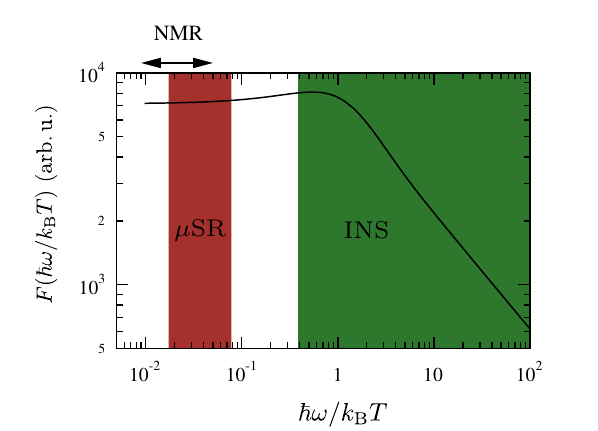}
\caption{\label{fig:Escales}(colour online). Comparison of the energy scales probed by the \musr\ experiment reported in this paper, the NMR experiment~\cite{Jeong2013} and the INS experiment~\cite{Povarov2015}. The solid line gives the scaling function~\cite{supplemental} $F(\hbar\omega/k_{\rm B}T)$ for the local dynamic structure factor with $K=1.1$.}
\end{figure}

Finally, in an inelastic neutron scattering experiment, the dynamic structure factor

\begin{equation}
\mathcal{S}^{\rho \rho}(\mathbf{q},\omega)=\int_{-\infty}^{\infty} \braket{S^\rho(\mathbf{0},0) S^\rho(\mathbf{r},t)} \exp[i(\omega t-\mathbf{q}\cdot\mathbf{r})] d{\bf r} dt
\label{eqn:dsf}
\end{equation}
can be probed as a function of energy and momentum transfer ($\rho=x,y,z$). INS therefore allows a certain region of interest in energy-momentum space to be selected and, in particular, it provides access to energy as an additional independent parameter. By comparing the experimental INS spectra with DMRG calculations it was possible to identify the parts of the spectrum where the transverse correlations described in Eq.~\ref{eqn:DSF} dominate~\cite{Povarov2015}. By using energy transfer and temperature as independent parameters it was possible to probe universal scaling over more than two decades in $\hbar\omega/k_{\rm B}T$, leading to an excellent agreement with DMRG calculations. Scaling was also observed by INS within the LRO phase $T<T_{\rm c}$ since the considered energy scales $0.1{\rm ~meV}<\hbar\omega<0.5{\rm ~meV}$ are well above the energy scale of the 3D interactions. The exact value of the determined Luttinger parameter at $\mu_0H=9$~T depends on the details of the analysis with values in the range: $K=1.25$ in Ref.~\onlinecite{Povarov2015}, $K=1.2(1)$ and $K=1.19(2)$ in Ref~\onlinecite{Schmidiger_thesis}.

In Fig.~\ref{fig:Escales} we compare the different scales in $\hbar\omega/k_{\rm B}T$ probed by \musr, NMR, and INS. \musr\ fills a gap that is inaccessible to both NMR and INS. In NMR similar energy scales could only be achieved at much higher fields implying that a different region of the phase diagram is being investigated. In INS such energy scales are practically inaccessible even with state-of-the-art cold neutron spectrometers.

In conclusion, using an exceptionally clean and well-characterized model system of a Tomonaga-Luttinger liquid, we have demonstrated that high-field \musr\ can be used to probe quantum-critical spin dynamics in a magnetic field range that is of great experimental interest in many topical materials. \musr\ fills a gap in energy scales that is inaccessible to established techniques such as NMR and INS. Model systems where \musr\ would prove particularly useful are those where INS is difficult to perform or universal behaviour at the lowest energy scales is of interest.

\begin{acknowledgments}
J.S.M.\ is grateful to Prof.\ Andrey Zheludev, Dr. Martin Klanj\ifmmode \check{s}\else \v{s}\fi{}ek, Dr.~Matthias Thede, and Dr.~Kirill Povarov for many helpful discussions. We thank Dr.\ James Lord for helpful comments concerning the detector deadtime corrections. J.S.M.\  is grateful for support by the ETH Zurich Postdoctoral Fellowship Program which has received funding from the European Union's Seventh Framework Programme for research, technological development and demonstration under grant agreement 246543. This project was supported by EPSRC (UK). Part of this work was carried out at the STFC ISIS facility and we are grateful for the generous provision of beam time. 
\end{acknowledgments}

%


\renewcommand*{\citenumfont}[1]{S#1}
\renewcommand*{\bibnumfmt}[1]{[S#1]}

\renewcommand{\thefigure}{S\arabic{figure}}
\renewcommand{\theequation}{S\arabic{equation}}

\section{Supplemental Material}
\subsection{Scaling function}
The full-scaling function for the transverse correlations in the dynamic structure factor is given by~\cite{Schulz1986,Sachdev1994,Chitra1997}:
\begin{equation}
    \begin{aligned}
&\mathcal{S}^{\perp \perp}\left ( \omega,q_{\parallel} \right) \propto  T^{1/2K -2} \, \times \\
& {\rm Im}\left\{ \left[ 1-\exp\left(-\frac{\hbar \omega}{k_{\rm B}T}\right)\right]^{-1} \Phi\left(\frac{\hbar \omega}{k_{\rm B}T},\frac{ u(q_{\parallel}-\pi)}{k_{\rm B}T}\right)\right\},
\end{aligned}
\label{eqn:Sfull}
\end{equation}
where $q_{\parallel}={\bf Q} \cdot {\bf a}$, $u$ is the field-dependent spin-wave velocity, and

\begin{equation}
\Phi(x,y)=\frac{\Gamma(\frac{1}{8K}-i\frac{x-y}{4\pi})}{\Gamma(1-\frac{1}{8K}-i\frac{x-y}{4\pi})}\frac{\Gamma(\frac{1}{8K}-i\frac{x+y}{4\pi})}{\Gamma(1-\frac{1}{8K}-i\frac{x+y}{4\pi})}.
\end{equation}
$\Gamma(x)$ is the complex $\Gamma$-function and $K$ is the Luttinger parameter. By integration of Eq.\ref{eqn:Sfull} with $y=\frac{ u(q_{\parallel}-\pi)}{k_{\rm B}T}$ it is therefore obvious that the transverse \emph{local} spin-spin correlations must obey the following scaling form
\begin{equation}
    \begin{aligned}
&\mathcal{S}^{\perp \perp}\left ( \omega \right) \propto  T^{1/2K-1} \, \times F\left(\frac{\hbar \omega}{k_{\rm B}T}\right).
\end{aligned}
\end{equation}
Numerical values were obtained by numerical integration of Eq.\ref{eqn:Sfull} over $q_{\parallel}$. 

\begin{figure}[htb]
\includegraphics[width=0.5\textwidth]{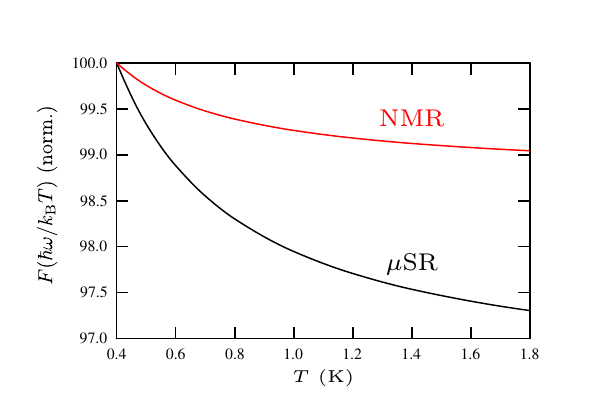}
\caption{\label{fig:Tdep}(color online). Temperature-dependence of the scaling function $F(\hbar\omega/k_{\rm B} T)$ for $K=1.1$ for both a \musr\ experiment at 4.8~T and a proton NMR experiment at 5~T. $F(\hbar\omega/k_{\rm B} T)$ is normalized to 100 at 0.4~K. Note that the normalization factor is  different for the \musr\ and the NMR curve as the probing energy scale is different: $F(2.7~\mu{\rm eV}/k_{\rm B} \times0.4 {\rm ~K})/F(0.88~\mu{\rm eV}/k_{\rm B} \times0.4 {\rm ~K})=1.024$.}
\end{figure}

Note that for the scaling properties of the local transverse correlation function $F(\hbar\omega/k_{\rm B} T)$ the value of the velocity $u$ is irrelevant. Even though the probing energy scales for a \musr\ and an NMR experiment are small, the non-zero value of $\hbar\omega$ ($\hbar\omega=2.7~\mu$eV for \musr\ at 4.8 T and $\hbar\omega=0.88~\mu$eV for proton NMR at 5~T) implies that \emph{there is a temperature-dependence} of $F(\hbar\omega/k_{\rm B} T)$ in the experimentally relevant temperature range. The temperature dependence of $F(\hbar\omega/k_{\rm B} T)$ is shown for a \musr\ and proton NMR in Fig.~\ref{fig:Tdep}. At a given field, the temperature dependence is stronger for a \musr\ experiment than for a proton NMR experiment by approximately the ratio of muon and proton magnetic moments.

\subsection{Field-dependent background}
The purpose of this section is to illustrate in more detail the technical challenges that were encountered during this experiment. Since the use of \musr\ in high longitudinal fields to study co-operative effects in magnetism is still in its infancy, we hope that this additional information will guide future high-LF \musr\ studies and help to avoid some of the technical issues.

\begin{figure}[htb]
\includegraphics[width=0.5\textwidth]{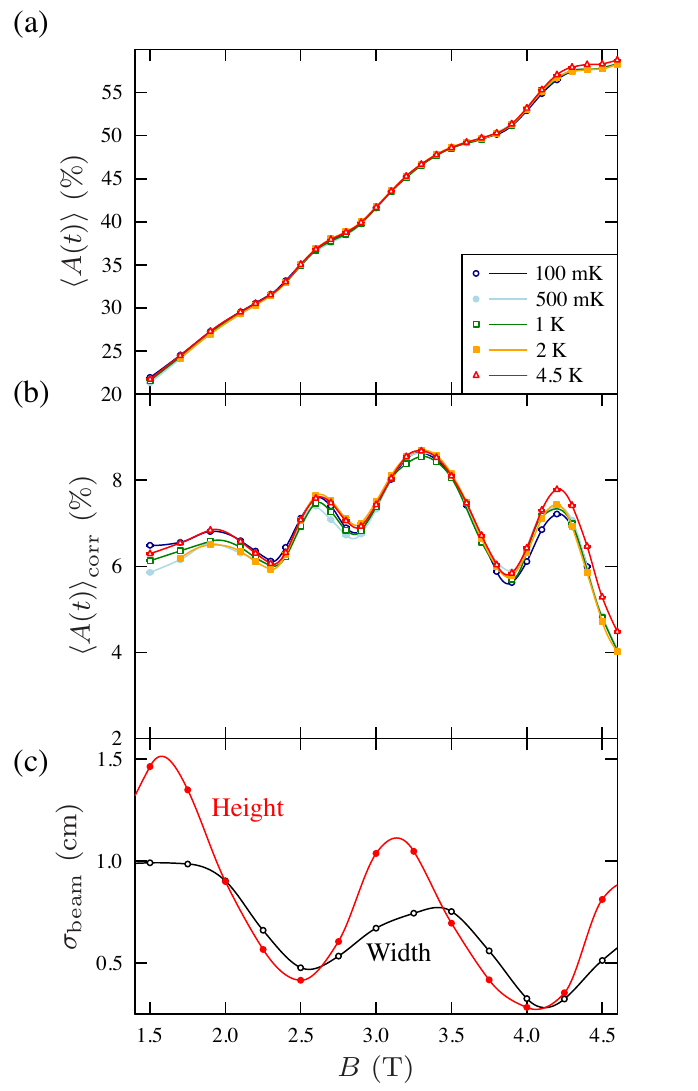}
\caption{\label{fig:Bdep}(color online). Field dependence of (a) the time-averaged asymmetry $\left<A (t)\right>$ (arbitrary offset). (b) Corrected time-averaged asymmetry $\left<A (t)\right>_{\rm corr}$. (c) Gaussian height and width of the muon beam~\cite{Lord_hifi}. }
\end{figure}

For the analysis in the main text, the experimental asymmetry was fitted with
\begin{equation}
A(t)=A_{\rm rel} \exp{(-t/T_1)} + A_{\rm nr},
\label{eqn:dimpy_relaxfit}
\end{equation}
where the relaxing asymmetry $A_{\rm rel}$ was kept fixed and $A_{\rm nr}$ is a field-dependent non-relaxing component. In addition to the temperature scans presented in the main text, we have performed extensive \emph{field scans} to investigate whether it is possible to probe the cross-over from the gapped quantum-disordered phase into the gapless TLL regime. For this purpose, a mosaic of single crystals was mounted on a silver backing plate which was attached to the cold finger of a dilution refrigerator. 

The asymmetry was determined from 
\begin{equation}
A(t)=\frac{N_{\mathrm{F}}(t)-\alpha_{\mathrm{exp}} N_{\mathrm{B}}(t)}
{N_{\mathrm{F}}(t)+\alpha_{\mathrm{exp}} N_{\mathrm{B}}(t)} \, ,
\end{equation}
where $\alpha_{\rm{exp}}$ is an experimental calibration constant accounting for different detector efficiencies, and $N_{\rm F}$ and $N_{\rm B}$ are the positron counts detected in a set of forward and backward detectors. $\alpha_{\mathrm{exp}}$ can normally be calibrated using an applied transverse field. However, in high-longitudinal fields $\alpha$ cannot be calibrated straightforwardly and was therefore set to $\alpha_{\mathrm{exp}}=1$. Effectively this leads to an arbitrary offset of the data.

In order to allow a model-independent analysis of the field-scan data, we have plotted the time-averaged asymmetry $\left<A (t)\right>$ for a range of temperatures in Fig.~\ref{fig:Bdep} (a). A suppressed value indicates a depolarization of the muon beam by temporal fluctuations. Since $A_{\rm rel}\ll A_{\rm nr}$, $\left<A (t)\right> \sim A_{\rm nr}$ (used for the temperature scans in the main text). $\left<A (t)\right>$ has an upwards slope combined with pronounced dips around 2.4~T, 2.9~T, and 3.9~T. The dips are more evident when a straight-line fit is subtracted from $\left<A (t)\right>$, which is shown as $\left<A (t)\right>_{\rm corr}$ in Fig.~\ref{fig:Bdep} (b). Also shown is the Gaussian width and height of the muon beam~\cite{Lord_hifi}. It is evident that the dips in $\left<A (t)\right>_{\rm corr}$ and $\left<A (t)\right>$ roughly coincide with the minima in the muon beam spot. This correspondence is not exact which may be related to the fact that the elliptical muon beam not only changes its size but also rotates as a function of field~\cite{Lord_hifi}. This suggests that their origin is due to the muon beam being focussed onto the sample, which only partly covers the silver backing plate. The sample depolarizes the muon beam more than the silver backing plate, at least at 2.4~T and above. While this is entirely consistent with our conclusion from the main text that we are indeed probing electronic dynamics in the sample, the complicated convolution of intrinsic dynamics with muon beam optics prevents more specific conclusions from the field scan data. The upward slope of $\left<A (t)\right>$ can be empirically attributed to a change in detector balance as a function of field.

Following these extensive measurements, we have therefore concluded that to avoid the above-mentioned issues: (i) A very large sample that fully covers the silver backing plate should be used to prevent spurious signals due to varying illumination of the sample by the muon beam. (ii) Temperature scans at fixed fields should be employed as these scans do not suffer from variation in the muon beam spot size. We note that for the present problem of finite temperature scaling in DIMPY these are also scientifically more apposite. (iii) The small relaxing amplitudes require high-statistics runs to be taken. (iv) Spurious relaxation due to detector deadtimes can potentially spoil the analysis of any small amplitude relaxation. Great care was taken to accurately correct for detector deadtimes by using silver background scans performed at a range of fields fully covering the field range studied here.

\end{document}